\documentclass[journal,twocolumn]{IEEEtran}

\usepackage{amsfonts}
\usepackage{times}
\usepackage{graphicx}
\usepackage{latexsym}
\usepackage{dsfont}
\usepackage{amssymb}
\usepackage{amsmath}
\usepackage{cite}
\usepackage{verbatim}
\usepackage{subfigure}

\def\bb0{{\mathbb{0}}}

\def\bb{{\mathbf{b}}}

\def\b0{{\mathbf{0}}}

\def\bS{{\mathbf{S}}}

\def\bX{{\mathbf{X}}}

\def\sf0{{\mathsf{0}}}

\usepackage{epstopdf}
\usepackage{enumerate}
\usepackage{algorithmicx}
\usepackage{algorithm}
\usepackage{amsmath}
\usepackage[noend]{algpseudocode}
\usepackage{float}
\usepackage{hyperref}
\usepackage{color}
\usepackage{makeidx}
\usepackage{bbm}
\usepackage{graphicx}
\usepackage{lipsum}
\usepackage{subfigure}
\usepackage{tablefootnote}
\usepackage{multirow}
\usepackage{multicol}
\usepackage{balance}
\usepackage{booktabs}
\usepackage{soul}
\usepackage[normalem]{ulem}
\usepackage{xcolor}

\newcommand{\comm}[1]{}

\begin{document}

\title{Multi-Modal Beam Prediction Challenge 2022: Towards Generalization}
\author{Gouranga Charan$^{1}$, Umut Demirhan$^{1}$, João Morais$^{1}$, Arash Behboodi$^{2}$, Hamed Pezeshki$^{3}$, and  Ahmed Alkhateeb$^{1}$\\ $^{1}$Wireless Intelligence Lab, Arizona State University, USA  \\ $^{2}$Qualcomm Technologies Netherlands B.V., Qualcomm AI Research\\ $^{3}$Qualcomm Technologies, Inc.  \thanks{Gouranga Charan, Umut Demirhan, João Morais, and Ahmed Alkhateeb are with the School of Electrical, Computer, and Energy Engineering, Arizona State University - Emails: \{gcharan, udemirhan, joao, alkhateeb\}@asu.edu. Arash Behboodi is with Qualcomm Technologies Netherlands B.V. and Qualcomm AI Research which is an initiative of Qualcomm Technologies, Inc. and/or its subsidiaries - Email: behboodi@qti.qualcomm.com. Hamed Pezeshki is with Qualcomm Technologies, Inc. - Email: hamedp@qti.qualcomm.com.}}

\maketitle

\begin{abstract}

Beam management is a challenging task for millimeter wave (mmWave) and sub-terahertz communication systems, especially in scenarios with highly-mobile users.  Leveraging external sensing modalities such as vision, LiDAR, radar, position, or a combination of them, to address this beam management challenge has recently attracted increasing interest from both academia and industry. This is mainly motivated by the dependency of the beam direction decision on the user location and the geometry of the surrounding environment---information that can be acquired from the sensory data. To realize the promised beam management gains, such as the significant reduction in beam alignment overhead,  in practice, however, these solutions need to account for important aspects. For example, these multi-modal sensing aided beam selection approaches should be able to generalize their learning to unseen scenarios and should be able to operate in realistic dense deployments.  

The ``Multi-Modal Beam Prediction Challenge 2022: Towards Generalization" competition is offered to provide a platform for investigating these critical questions. In order to facilitate the generalizability study, the competition offers a large-scale multi-modal dataset\footnote{Dataset was created and evaluated by the ASU researchers.} with co-existing communication and sensing data collected across multiple real-world locations and different times of the day. In this paper, along with the detailed descriptions of the problem statement and the development dataset, we provide a baseline solution that utilizes the user position data to predict the optimal beam indices. The objective of this challenge is to go beyond a simple feasibility study and enable necessary research in this direction, paving the way towards generalizable multi-modal sensing-aided beam management for real-world future communication systems.

\end{abstract}


\section{Introduction} \label{sec:Intro}

Current and future communication systems are moving to higher frequency bands --- mmWave in 5G and potentially sub-terahertz in 6G and beyond. The large available bandwidth at these high frequency bands enables the communication systems to satisfy the increasing data rate demands of the emerging applications, such as autonomous driving, 8K video streaming, and immersive mixed-reality \cite{Rappaport2019}. These systems, however, are required to deploy large antenna arrays at the transmitters and/or receivers and use narrow directive beams to guarantee sufficient receiver power. Selecting the best beams out of a large pre-defined codebook is typically associated with high beam training overhead. This high beam training overhead makes it difficult for these systems to obtain frequent (accurate) beams, thereby making it challenging to support highly-mobile and latency-sensitive applications.

Developing solutions for the mmWave beam training and channel estimation overhead has attracted considerable interest over the last decade \cite{Zhang2021a,Alkhateeb2014,Jayaprakasam2017,HeathJr2016}. The solutions have generally focused on: (i) constructing adaptive beam codebooks \cite{Zhang2021a,Zhang2021b}, (ii) designing beam tracking techniques \cite{Jayaprakasam2017}, and (iii) leveraging the channel sparsity and efficient compressive sensing tools \cite{Alkhateeb2014,HeathJr2016}. The classical approaches, however, can typically save only one order of magnitude in the training overhead, which is not sufficient for the systems with very large antenna arrays, especially when serving highly-mobile and latency critical applications. 

The challenges faced by the classical approaches motivated the development of machine learning (ML) solutions \cite{Alkhateeb2018, Rezaie2022, Morais2022a, Sub6PredMmWave, Muhammad2020a,  Charan2022a, Charan2022b, Jiang2022b, Jiang2022a, Demirhan2022a } that leverage prior observation and side information for fast mmWave/THz beam prediction. The dependency of mmWave/ THz communication systems on the LOS links between the transmitter and receiver means that the awareness of their locations and the surrounding environment (geometry of the buildings, moving scatterers, etc.) could potentially help the beam selection process. To that end, the overall sensing information of the environment could be utilized to guide the beam management process and reduce beam training overhead significantly. For example, sensory data such as user position/orientation \cite{Rezaie2022, Morais2022a,Khan2020a}, sub-6GHz channel information \cite{Sub6PredMmWave}, RGB images \cite{ Muhammad2020a,  Charan2022a, Charan2022b,Jiang2022b}, LiDAR point clouds \cite{Jiang2022a}, and radar measurements \cite{Demirhan2022a} can enable the transmitter/receiver to decide on where to point their beams or at least narrow down the candidate beam steering directions. 


\begin{figure*}[!t]
	\centering
	\includegraphics[width=0.95\linewidth]{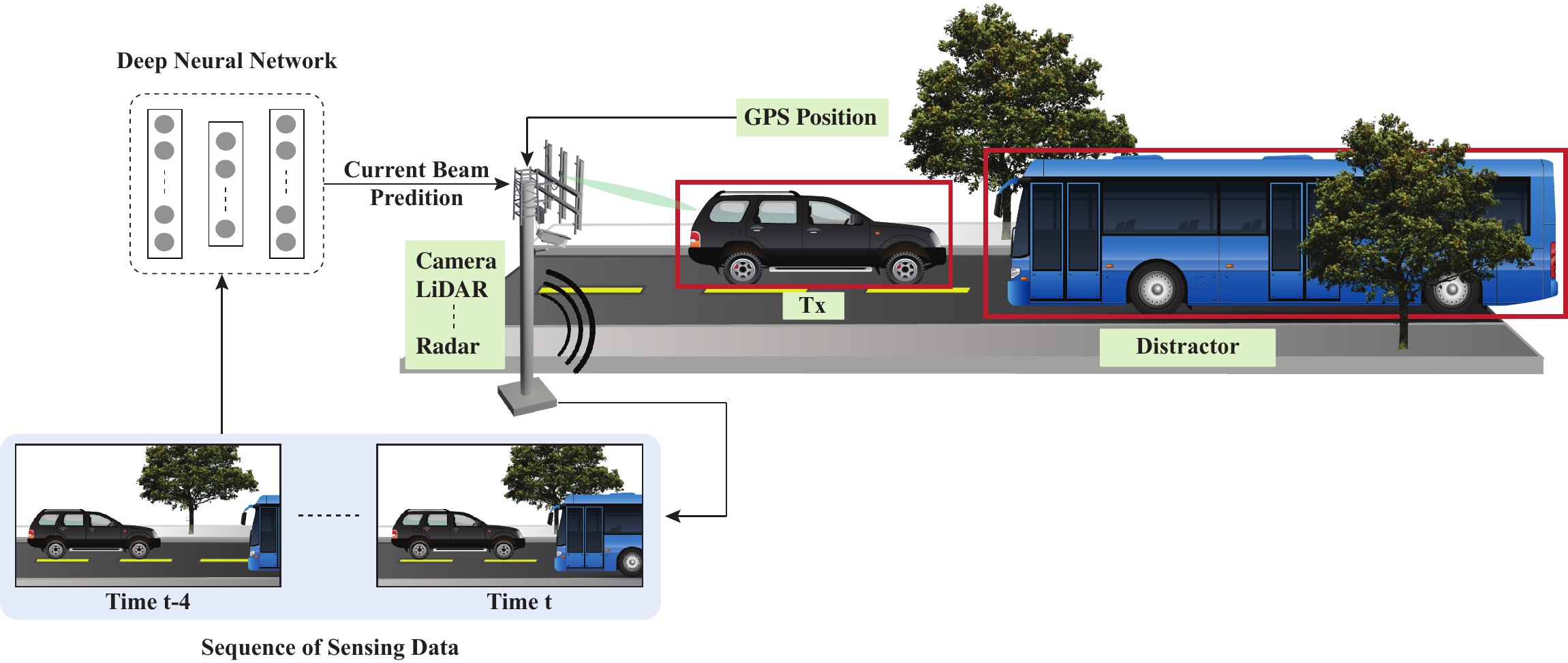}
	\caption{This figure illustrates the considered system and highlights the value of leveraging the side information (user position, vision, LiDAR and radar) for efficient mmWave/Thz beam prediction.}
	\label{fig:main_fig}
\end{figure*}


Recent work on sensing-aided beam prediction has shown initial promising results in utilizing sensory data such as GPS positions, RGB images, LiDAR point clouds, and radar observations for the beam prediction problem. However, these solutions may be limited due to the following reasons: (i) These approaches were usually developed and evaluated based on synthetic datasets and might experience lower performance in real-world scenarios. (ii) Most of these solutions are trained and tested on the same dataset and may lack capabilities to generalize across different scenarios/locations. (iii) Furthermore, each of these sensory modalities, has its limitations. For example, practical positioning sensors do not normally provide accurate enough positions for narrow beam alignment, and visual/camera data may be sensitive to lighting/weather conditions. Such limitations will affect the performance of the proposed solutions in real-world scenarios. Therefore, to overcome these limitations and take sensing-aided communication ideas one step closer to real-world deployments, we need to answer the following important questions: (i) \textbf{Can sensing-aided beam prediction solutions perform well on real-world data?} (ii) \textbf{How to utilize cross-modality information and efficiently fuse different modalities?} and (iii) \textbf{Can the developed ML models that are trained on certain scenarios generalize their beam prediction/learning to new scenarios that they have not seen before?}

\section{Multi-Modal Beam Prediction: \\ System Model and Problem Formulation} \label{sec:sys_model_prob_form}
This ML challenge targets addressing the important question of large beam training overhead in mmWave/THz communication systems. In the challenge, the participants are asked to design machine learning-based solutions that can be trained on a dataset of a few scenarios and then generalize successfully to the other scenarios that the models have not seen before. Therefore, given a multi-modal training dataset consisting of data collected at different locations with diverse environmental features, the objective is to develop machine learning-based models that can adapt to an entirely new location and perform accurate sensing-aided beam prediction at an entirely new location that is not a part of the training dataset. In particular, the beam prediction task is posed as a multi-modal classification problem: The objective is to predict the best beamforming vector from a pre-defined codebook by utilizing the sensing data available from the wireless environment. In this section, we first describe the adopted wireless communication system model. Then, we formulate the sensing-aided beam prediction problem targeted in this challenge.

\subsection{System Model} \label{subsec:sys_model}
In this work, we adopt the system model illustrated in Fig.~\ref{fig:main_fig}, where a mmWave basestation equipped with an $M$-element uniform linear array (ULA) and a suite of sensors (RGB camera, 3D LiDAR and radar) is serving a mobile user. The user is equipped with a single-antenna transmitter and a GPS receiver capable of collecting real-time location. The adopted communication system employs OFDM transmission with K subcarriers. To serve the mobile user, the basestation is assumed to employ a pre-defined beamforming codebook $\boldsymbol{\mathcal F}=\{\mathbf f_q\}_{q=1}^{Q}$, where $\mathbf{f}_q \in \mathbb C^{M\times 1}$ and $Q$ is the total number of beamforming vectors. In the downlink transmission, if $\mathbf h_{k}[t] \in \mathbb C^{M\times 1}$ denotes the channel between the basestation and the user at the $k$-th subcarrier and time $t$, then the received signal at the basestation can be written as 
\begin{equation}\label{eq:sys_mod}
	y_{k}[t] = \mathbf h_{k}^T[t] \mathbf f_{q[t]}x + v_k[t],
\end{equation}
where $\mathbf f_{q[t]} \in \boldsymbol{\mathcal F}$ is the beamforming vector applied at time $t$ and $v_k[t]$ is the receiver noise with a complex Gaussian distribution $\mathcal N_\mathbb C(0,\sigma^2)$. The transmitted complex symbol $x\in \mathbb C$ needs to satisfy the average power constraint  $\mathbb E\left[ |x|^2 \right] = P$, where $P$ is the average symbol power.


\begin{figure*}[!t]
	\centering
	\includegraphics[width=0.95\linewidth]{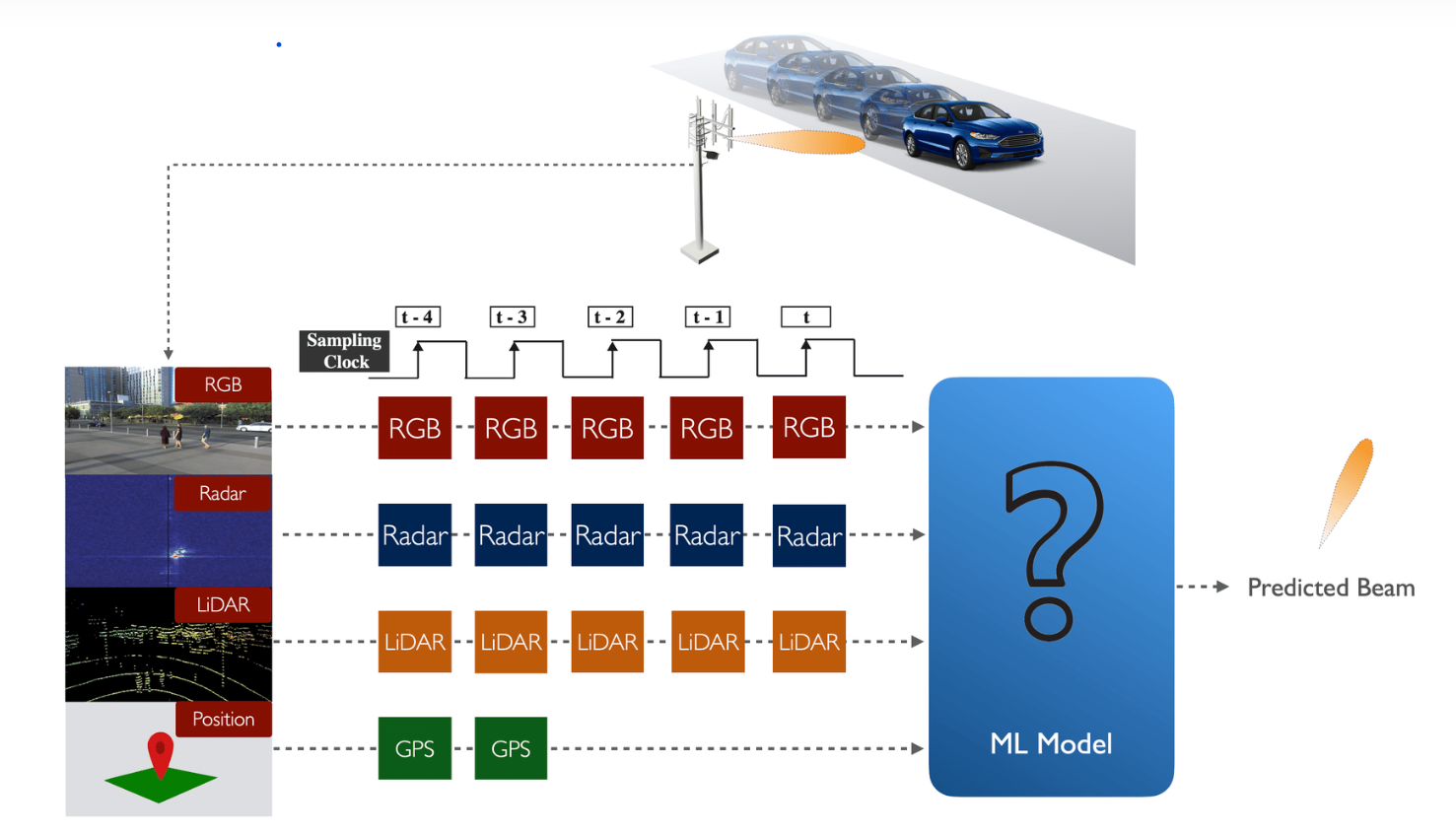}
	\caption{This figure illustrates the schematic representation of the input data sequence utilized in this challenge tasks. Each data sample comprises of a sequence of $5$ sensory data (RGB images, LiDAR point cloud and radar data). We also provide the ground-truth GPS locations for the first two instance in the sequence.}
	\label{fig:sequence_data}
	\vspace{-2mm}
\end{figure*}


\subsection{Problem Formulation}\label{subsec:prob_form}

Given the system model presented in Section~\ref{subsec:sys_model}, the task of beam prediction entails determining the index of the optimal beamforming vector ${q^{\star}[t]}$ out of the indices of the candidate beams in the codebook $\{1, \ldots, Q\}$, such that the average beamforming gain is maximized. Mathematically, the beam selection problem can be expressed as
\begin{equation}\label{eq:beam_training}
	q^{\star}[t] = \underset{q \in \mathcal \{1, \ldots, Q\}}{\text{argmax}} \frac{1}{K}\sum_{k=1}^{K} |\mathbf h_{k}^T[t] \mathbf f_q |^2.
\end{equation}

Conventionally, the optimal beam index is computed by (i) utilizing the explicit channel information, which, in general, is hard to acquire in mmWave/THz systems, or (ii) performing exhaustive search over the beam codebook, which is typically associated with large beam training overhead for the mmWave/THz systems with large antenna arrays. These challenges make it difficult for these high-frequency systems to support high-mobility and latency critical applications.

To address the large beam training overhead in mmWave/THz communication systems, the primary track in this challenge is  the multi-modal sensing-aided beam prediction problem. The challenge task entails observing a sequence of sensory data captured at the basestation to predict the current optimal beam indices from a pre-defined codebook. Along with the sequence of the sensing data (i.e., image, LiDAR and radar), we also provide the ground-truth GPS measurements (latitude and longitude) of the transmitter in the scene for a subset of the sequence. The primary reasoning behind providing partial positional information is that the GPS measurements are user side information and might not be available for every time instance. However, to predict the optimal beam indices for a scenario with multiple probable transmitting candidates, it is critical to first identify the transmitter among the different objects. To that end, the scenarios selected for this challenge are primarily multi-candidate, i.e., more than one object is present in the wireless environment. Here, one approach can be to first identify the transmitter in the scene, for which the positional data can be utilized. To formalize the available data, we define:
\begin{itemize}
	\item $\bX_P[t] \in \mathbb R^2$  as the two-dimensional position vector of the transmitter (consisting of the latitude and longitude information) at time step $t$.
	\item $\bX_I[t] \in \mathbb{R}^{W_I \times H_I \times C_I}$ as the corresponding RGB image, captured by a camera installed in the basestation at time $t$, where $W_I$, $H_I$, and $C_I$ are the width, height, and the number of color channels of the image.
	\item $\bX_L[t] \in \mathbb{R}^{D_L \times H_L \times W_L}$ as the 3D point cloud captured by the LiDAR installed in the basestation at time $t$, where $D_L$, $H_L$, and $W_L$ are the depth, height, and the width of the point cloud data, respectively.
	\item $\bX_R[t] \in \mathbb{R}^{M_R \times S_R \times A_R}$ as the radar measurements captured at time $t$, where $M_R$, $S_R$, and $A_R$ are the number of radar antennas, the number of samples per chirp, and the number of chirps per frame, respectively.
\end{itemize}

It is important to note here that instead of making the predictions based on a single sample, the objective of this challenge is to utilize a sequence of $r$ basestation-side current and previously observed sensory data for the RGB images, radar, and LiDAR, and $r'$ previously observed position data from the user-side, to predict the current optimal beam indices. At any time instant $\tau\in \mathbb Z$, this sequence can be written as
\begin{equation}
	{\bS}[\tau]=\left\lbrace  \left\lbrace  \bX_P[t'] \right\rbrace_{t' = \tau-r+1}^{ \tau - r + r' + 1} ,
	\left\lbrace  \bX_I[t], \bX_L[t], \bX_R[t]  \right\rbrace_{t = \tau-r+1}^{ t = \tau} \right\rbrace , 
\end{equation}
where $r, r' \in \mathbb Z$ is the lengths of the input sequences (i.e., the observation windows) to predict the current optimal beam indices. In this challenge, the values of $r$ and $r'$ are selected to be $5$ and $2$, respectively. Specifically, each data sample comprises of a sequence of $5$ basestation-side sensory data (RGB images, LiDAR point cloud and radar data) along with $2$ user-side sensory data (GPS locations) for the first two instances in the sequence. As shown in Fig.~\ref{fig:sequence_data}, at any time instant $t$, a sequence of 5 samples (current and previously observed sensing data, i.e., $\left\lbrace t-4, …, t\right\rbrace )$ is provided. We also provide the GPS locations of the user for the first two samples in the sequence $({t-4, t-3})$. 

Participants are expected to design a machine learning solution that takes this data sequence and learns to predict the optimal beam index at time $t$. The participants are free to utilize either all the data modalities or a subset of the data modalities. We formulate this problem in this way to mimic a real-world scenario where not all the modalities (especially position) are available at all the samples. In reality, the position information may be captured by the user and then transmitted back to the basestation for downstream applications. Hence, the position data might not be available for every time instance and may be subject to additional delays. Further, the sampling frequency of GPS sensors is $\approx10$Hz, which is less than the typical sampling rate for the other modalities. The objective of this challenge is to find a prediction/mapping function $f_{\Theta}$ of parameters $\Theta$, that utilizes the available sensory data $\mathcal{S}[\tau]$ to predict (estimate) the optimal beam index $\hat{q}[\tau] \in \{1, \ldots, Q\}$ with high fidelity. The mapping function can be formally expressed as
\begin{equation}
	f_{\Theta}: \mathcal{S}[\tau] \rightarrow  \hat{q}[\tau].
\end{equation}

\section{Testbed Description and Development Dataset}\label{sec:testbed_datset}

\begin{figure}[!t]
	\centering
	\includegraphics[width=1\linewidth]{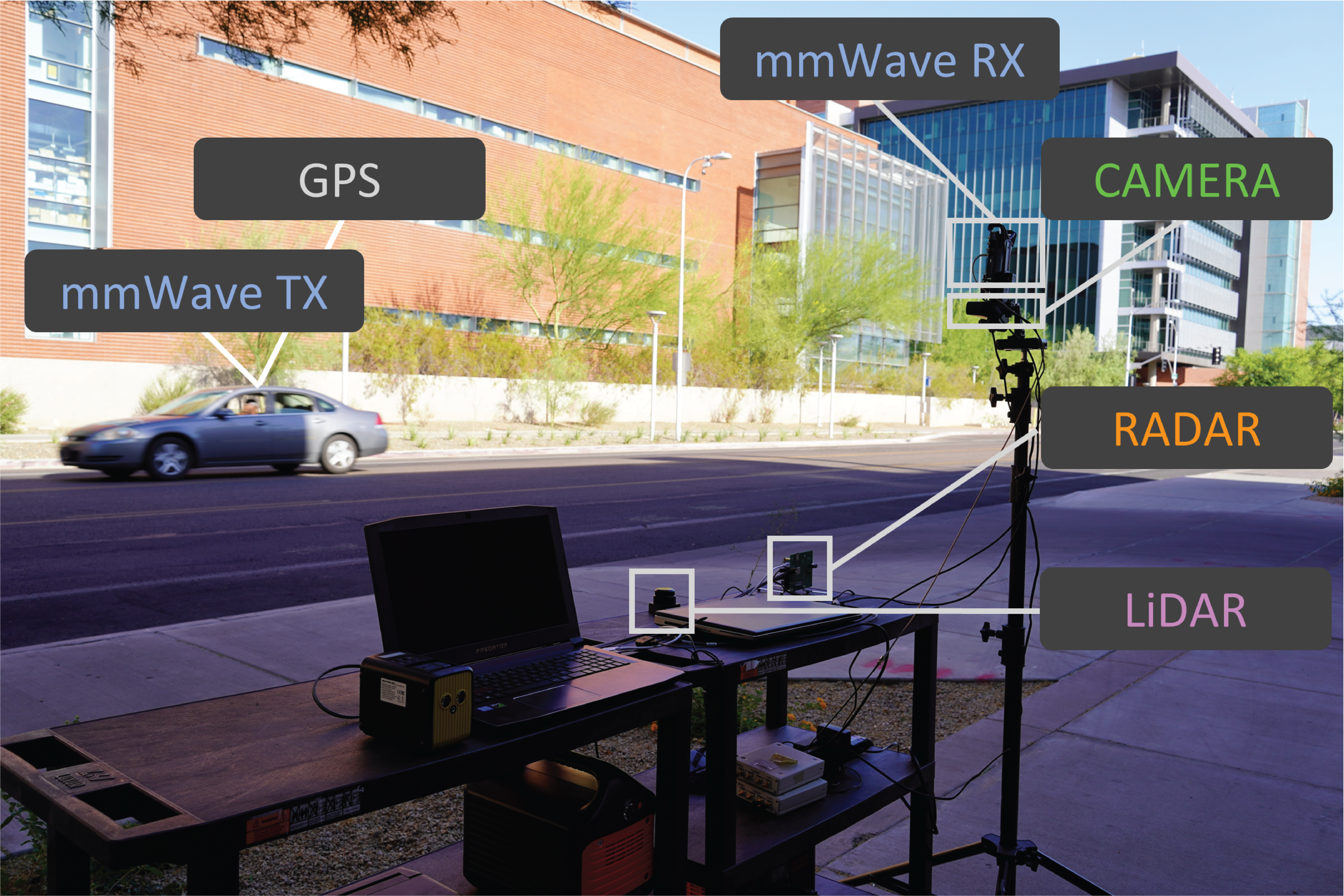}
	\caption{This figure presents the DeepSense 6G testbed utilized to collect the multi-modal data for these scenarios.}
	\label{fig:testbed}
\end{figure}

Developing efficient solutions for sensing-aided beam prediction and accurately evaluating them requires the availability of a large-scale real-world dataset. With this motivation, we built the DeepSense 6G dataset, the first large-scale real-world multi-modal dataset with co-existing communication and sensing data. Fig.~\ref{fig:testbed} highlights the DeepSense 6G testbed and the overall data collection and post-processing setup. In this section, we first present an overview of the data collection testbed. Then, we present the development dataset adopted for this ML challenge. 

\begin{figure}[!t]
	\centering
	\includegraphics[width=0.85\linewidth]{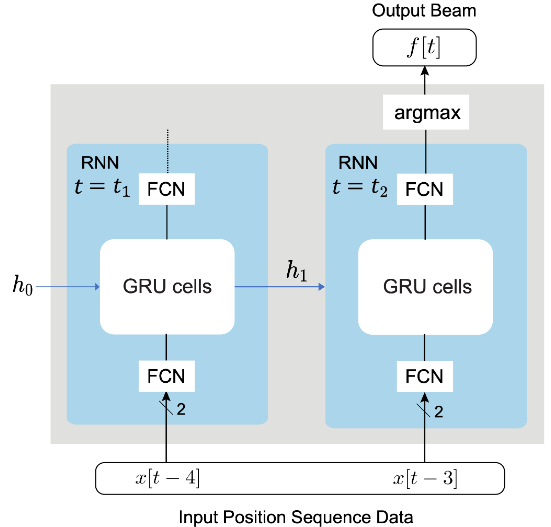}
	\caption{This figure presents the proposed position-aided future beam prediction solution. It comprises of a recurrent neural network that receives the sequence of position data as input and learns to predict the optimal beam indices at the 3rd future time instance.}
	\label{fig:baseline_sol}
\end{figure}


\subsection{Data Collection Testbed}  \label{subsec:testbed}
In this challenge, we build development/challenge datasets based on the DeepSense data from scenarios $31 - 34$. The DeepSense testbed $5$ is utilized for this data collection, which comprises of a stationary and a mobile unit.The mobile unit, acting as a transmitter, is equipped with a 60GHz quasi-omni antenna and a GPS receiver to collect the real-time location of the user. The stationary unit adopts a $16$-element ($M = 16$) 60GHz-band phased array. This unit receives the transmitted signal with an over-sampled codebook of $64$ pre-defined beams ($Q = 64$). Furthermore, the basestation is equipped with an RGB camera, a $3$D LiDAR and a frequency modulated continuous wave (FMCW) radar working at 77 GHz. The data collected at each time instant comprises of the GPS position of the user, RGB images, radar I/Q samples and LiDAR point-cloud data. The corresponding $64$ element power vectors are obtained by performing beam training with a $64$-beam codebook at the receiver (with omni-transmission at the transmitter). The LiDAR, radar and visual data are captured by the sensors installed at the basestation. The position data is collected from the GPS receiver installed on the mobile unit (vehicle). For further details, please refer to the data collection testbed description in the DeepSense 6G dataset \cite{DeepSense}.

\subsection{Development Dataset}\label{subsec:dev_dataset}

The primary goal of this challenge is to study the generalizability of the sensing-aided beam prediction solutions. With this motivation, we generated the final challenge training and test dataset. The development dataset consists of:
\begin{itemize}
	\item \textbf{Training Dataset:} The training dataset comprises the data from the three scenarios $32 - 34$
	\item \textbf{Test Dataset:} 50$\%$ of the data is from scenarios $32 - 34$ and the remaining 50$\%$ is from the unseen scenario $31$.
\end{itemize}
In this challenge, we have a single track, \textit{multi-modal beam prediction}. This track includes data from all modalities. However, participants can select a subset of these data modalities and design the ML solution based on them. In addition to the full-dataset comprising of all the modalities, we also provide the following dual modality sub-datasets for ease of access and development: (i) vision-position, (ii) LiDAR-position and (iii) radar-position. For example, in the vision-position beam prediction sub-dataset, we provide the following: 
\begin{itemize}
	\item \textbf{Training Dataset:} Each data sample comprises of the sequence of $5$ images and the ground-truth GPS locations of the transmitter for the first two instances in the sequence. We also provide the $64$ element power vector corresponding to the $5$-th sample in the sequence. The dataset also provides the optimal beam indices as the ground-truth labels. 
	\item \textbf{Test Dataset:} To motivate the development of efficient ML models, we provide a test dataset, consisting of the sequence of $5$ input RGB images and $2$ ground-truth GPS locations. The ground-truth labels are hidden from the users by design to promote a fair benchmarking process. 
\end{itemize}

\begin{figure}[!t]
	\centering
	\includegraphics[width=1\linewidth]{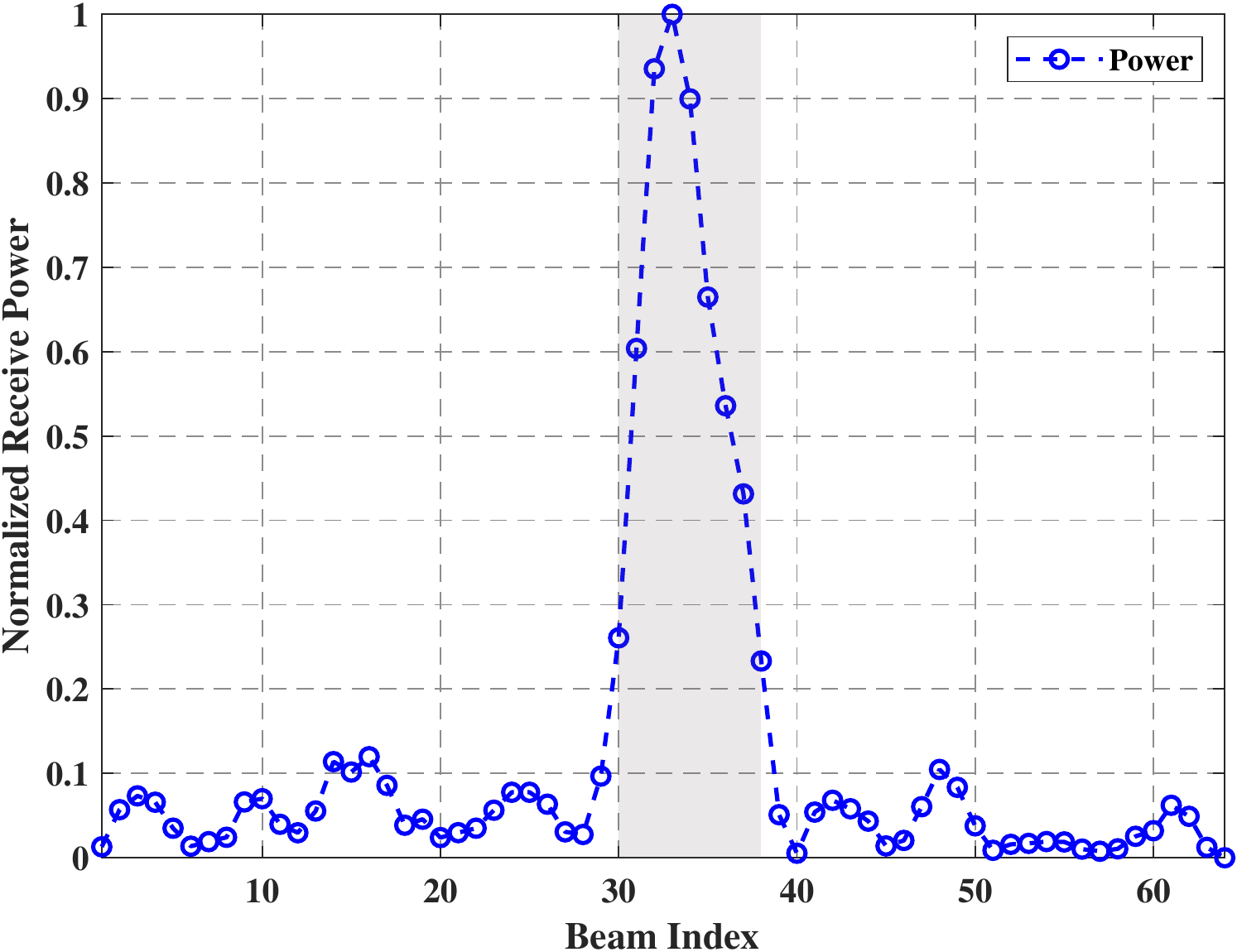}
	\caption{This figure plots the ground-truth receive power versus the beam indices for an example sample. The power corresponding to the beam indices in the highlighted region are closer to the maximum receive power.  }
	\label{fig:beam_vs_power}
\end{figure}


\section{Evaluation Metric} \label{sec:eval_metric}
To evaluate the proposed solutions, we provide the following metric named as the "Distance-based Accuracy Score (DBA)". For this, we utilize the top-$K$ predicted beams. The DBA-Score is defined as
\begin{equation}
	\text{DBA-Score} = \frac{1 }{K} \sum_{k=1}^{K} Y_k, 
\end{equation} 
where $Y_k$ is defined as
\begin{equation}
	Y_k = 1 - \frac{1}{N} \sum_{n=1}^{N} \underset{1\leq k' \leq k}{\min} \left[ \min \left( \frac{|\hat{y}_{n,k'} - y_n|}{\Delta}, 1 \right) \right],  
\end{equation}
where $y_n$ and $\hat{y}_{n,k'}$ are the ground-truth beam index and the predicted beam index of sample $n$, respectively. N is the total number of samples in the test set. To clarify the notation with an example, $\hat{y}_{n,2}$ represents the prediction for the second most-likely beam of sample $n$. $\Delta$ is a hyperparameter. In this challenge, we will be using $K = 3$ and $\Delta = 5$. The solution with the highest DBA-Score will be the winner. We present the intuition behind the proposed metric in the following section.

\section{Position-Aided Beam Prediction: \\ A Baseline Solution} \label{sec:baseline_sol} 

In the challenge dataset, a sequence of $5$ basestation-side samples are provided along with the user GPS locations for the first two instances. The participants are expected to predict the optimal beam indices corresponding to the $5$th data samples in the sequence. As a baseline solution, we utilize only the two user GPS locations provided in each data sample to predict the future beam at the $5$th time instance, i.e., the $3$rd future instance (as compared to the latest available GPS locations). In this section, we present the proposed machine learning-based solution and the beam prediction performance achieved by the baseline solution.

\subsection{Data Processing} \label{subsec:data_processing}
A position sample is two-dimensional, composed of geographical coordinates in decimal degrees, i.e., a latitude in $[-90^{\circ},90^{\circ}]$ and a longitude in $[-180^{\circ},180^{\circ}]$. Position data is normalized in three steps: (i) transform to cartesian XY coordinates using the UTM projection \cite{UTM,UTM_Wiki}; (ii) compute the differences between all the UE positions and the BS position; (iii) apply a min-max normalization on the position differences. Note that the min-max normalization of $N \times 2$ matrix is the same as applying the min-max to each column individually, i.e., compute the minimum and maximum values across of the column and for each point $x$ in the column, transform it to $x' = (x - \min) / (\max - \min)$. We, further, divide the available training dataset of the challenge into training, validation and test set following a $70-20-10\%$ split ratio. 

\begin{figure}[!t]
	\centering
	\includegraphics[width=1\linewidth]{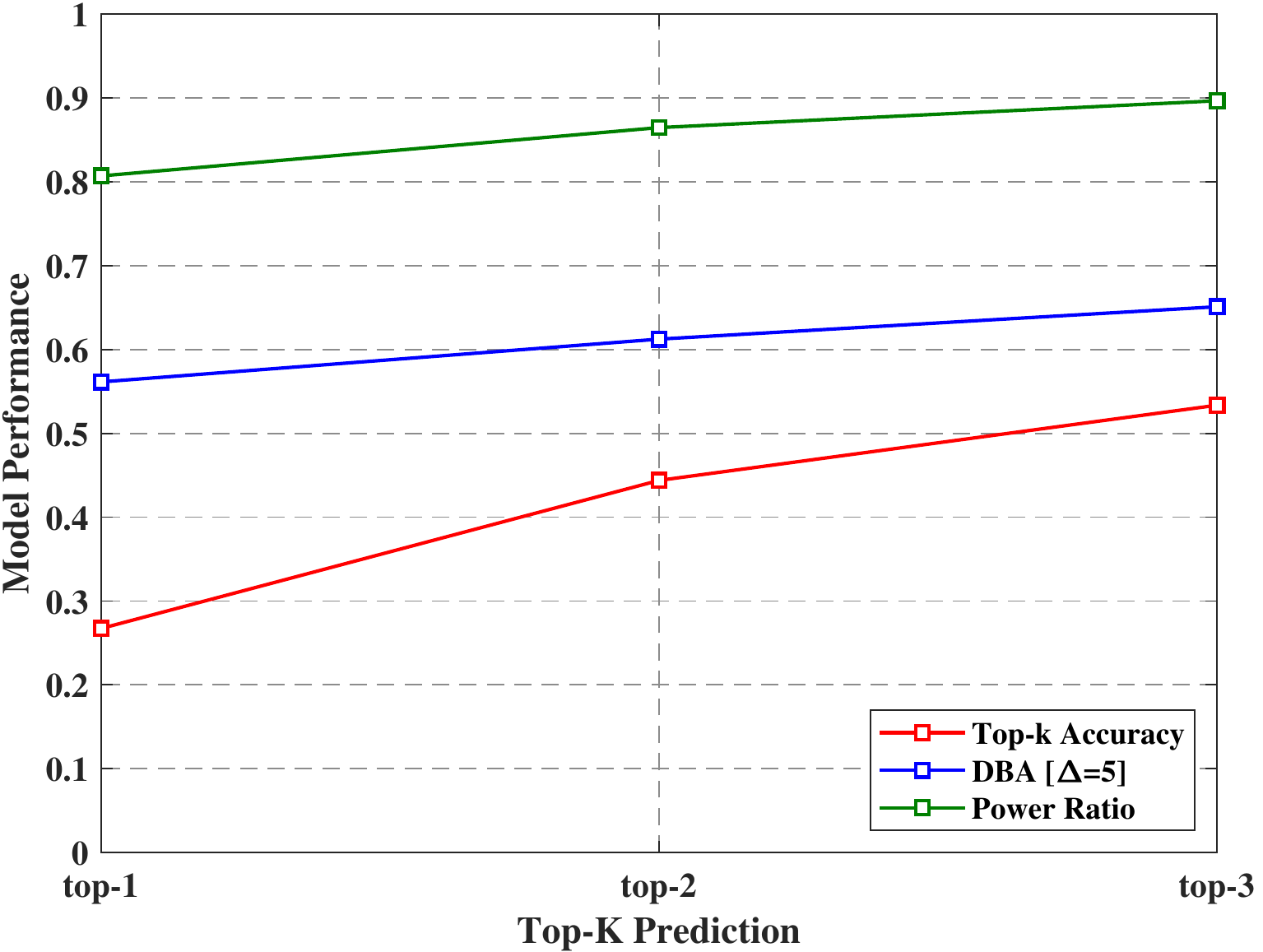}
	\caption{This figure presents the comparison among the three different metrics, i.e., top-$K$ accuracy, DBA score and the power ratio. It is observed that the DBA score has a higher correlation with the power ratio, which inherently mimics the model performance in real-world. }
	\label{fig:performance_comparison}
\end{figure}

\begin{table}[!t]
	\caption{Beam Prediction Performance }
	\centering
	\setlength{\tabcolsep}{5pt}
	\renewcommand{\arraystretch}{1.5}
	\begin{tabular}{@{}cccc@{}}
		\toprule
		\multicolumn{1}{c|}{\multirow{2}{*}{\textbf{Dataset}}} & \multicolumn{3}{c}{\textbf{Beam Prediction Accuracy}}                                     \\ \cmidrule(l){2-4} 
		\multicolumn{1}{c|}{}                                  & \multicolumn{1}{c|}{\textbf{top-1}} & \multicolumn{1}{c|}{\textbf{top-2}} & \textbf{top-3} \\ \midrule \midrule
		\multicolumn{1}{c|}{\textbf{Test}}                     & \multicolumn{1}{c|}{0.2672}          & \multicolumn{1}{c|}{0.4439}          & 0.5336          \\ \midrule 
		\multicolumn{4}{c}{\textbf{DBA-Score}}                                                                                                             \\ \midrule
		\multicolumn{1}{c|}{\textbf{Test}}                     & \multicolumn{1}{c|}{0.5614}         & \multicolumn{1}{c|}{0.6124}         & 0.6511         \\ \bottomrule
	\end{tabular}
	\label{tab:baseline_perf}
\end{table}


\subsection{Proposed Solution and Performance Evaluation} \label{subsec:prop_soln}
In this subsection, we first present the proposed solution followed by the beam prediction performance achieved by the ML model. Then, we provide some insights on why we adopted DBA-score as the preferred evaluation metric. 

\textbf{Proposed Solution:} The solution utilizes a recurrent neural network to predict the optimal beam indices. Specifically, we consider a two-stage gated recurrent unit (GRU), followed by a fully-connected layer acting as a classifier. More specifically, the model receives a sequence of two normalized positional data as input and predicts the $3$-rd future optimal beam indices. The proposed ML model is trained and validated on the task-specific dataset as presented in Section~\ref{subsec:dev_dataset}. The cross-entropy loss with the Adam optimizer is used to train the model.

\textbf{Performance Evaluation:} Table~\ref{tab:baseline_perf} presents the beam prediction performance achieved by the proposed solution with the metrics top-$K$ accuracy and DBA-score. The top-$K$ accuracy is defined as the percentage of the test samples whose ground-truth beam index lies in the K most likely predicted beams. From the table, it is observed that the proposed solution can successfully predict the $3$rd future-beam with $\approx 53\%$ top-$3$ accuracy on the test set. Furthermore, we calculate the DBA-Score achieved by the proposed solution. For this, we utilize the evaluation metric presented in Section~\ref{sec:eval_metric} and utilize the top-$3$ predicted beams, i.e., K = $3$. The proposed solution achieves a DBA-score of $0.65$ corresponding to the top-3 predicted beams. This beam prediction performance achieved by the baseline solution highlights the efficiency and potential of the proposed sensing-aided solution for the mmWave/THz beam prediction task. The question that still remains is \textit{why do we utilize DBA-score for ranking in this competition?} Next, we provide the intuition behind utilizing DBA-score as the preferred metric in this competition.

\textbf{Does top-$K$ accuracy capture the true model performance?} Most prior work for sensing-aided beam prediction \cite{Morais2022a, Charan2022a, Charan2022b, Jiang2022a, Demirhan2022a} have utilized the top-$K$ accuracy as the primary metric to evaluate the performance of the proposed solution. Similar to the classification problem in machine learning (for example, in object detection), the top-$K$ accuracy adopts a binary approach where a prediction is labeled as correct if the ground-truth beam lies in the predicted top-$K$ beams. Although suitable for a task like image classification, this approach might not be the best suited for a wireless communication task, such as the beam prediction task studied in this work. To provide further insight into this claim, refer to Fig.~\ref{fig:beam_vs_power}. This figure plots the receive power versus the beam indices for a particular sample. The optimal beam index in this example is $32$, i.e., the beam corresponding to the index of the maximum receive power. However, as shown in this figure, beams inside the highlighted region (indices $28 - 35$) may still result in sufficient receive power. This analysis highlights that even though the ground-truth beam might not lie within the predicted top-$K$ beams, the receive power can be significantly high. Therefore, a more holistic metric is needed to truly measure the beam prediction performance of a model. 


\textbf{Why DBA?} To answer this question, let us first define a metric that targets system performance with more rigor. The metric of choice is the average power ratio between the ground-truth and the predicted power, defined as 
\begin{equation}
	P_{R} = \frac{1}{N} \sum_{n=1}^{N} \left( \frac{P^{n}_{\hat{ q}} - P_{v}}{P^{n}_{q}- P_{v}} \right), 
\end{equation}
where $P_{v}$ is the noise power of the particular scenario, $P^{n}_{\hat{q}}$ and $P^{n}_{q}$ are the power of the predicted beam and the ground-truth beam of the sample $n$, respectively. The minimum power value in the measurement of each scenario is assigned as the noise floor (power) for that particular scenario. For the cases with $K > 1$, we compute the receive power corresponding to each of the $K$ predicted beams and assign the maximum value among the $K$ receive power as the predicted power. The average power ratio ($P_R$), which computes the ratio between the predicted and ground-truth receive power, provides a more realistic measure of the model performance; it indictes the power gain with the beam selection. However, the ground-truth receive power corresponding to each beam index might not be available in practice, making it difficult to utilize this power ratio as a universal machine learning evaluation metric.

With this motivation, in this work, we propose to develop the DBA performance evaluation metric, as presented in Section~\ref{sec:eval_metric}. Instead of following a hard binary approach as the previously employed top-$K$ accuracy, the DBA score adopts a softer evaluation approach. It essentially computes the distance of the predicted beams from the ground-truth beams and assigns scores based on how far the predicted beam is from the ground-truth beam. In Figure~\ref{fig:performance_comparison}, we compare the three metrics, i.e., top-$K$ accuracy, DBA score, and the power ratio. The DBA score has higher correlation with the power ratio metric compared to the top-$K$ accuracy, and hence better captures the real-world performance. These properties motivates adopting the DBA-score for the DeepSense beam prediction machine learning competition.

\balance

\end{document}